# Revealing Material-Dependent Bicircular High-Order Harmonic Generation in 2D Semiconductors via Real-Space Trajectories

Qing-Guo Fan[1,2], Kang Lai[1], Wen-hao Liu[1], Zhi Wang[1], Lin-Wang Wang[1,2], Jun-Wei Luo[1,2]

[1] State Key Laboratory of Semiconductor Physics and Chip Technologies, Institute of Semiconductors, Chinese Academy of Sciences, Beijing 100083, China

[2] Center of Material Science and Optoelectronics Engineering, University of Chinese Academy of Sciences, Beijing 100049, China

**Abstract:** Solid-state high-order harmonic generation (HHG) presents unique features different from gases. Whereas the gaseous harmonics driven by counter-rotating bicircular (CRB) pulse universally peak at a "magic" field ratio $E_{2\omega}:E_\omega \approx 1.5:1$, crystals exhibit significant material-dependent responses. In monolayer $MoS_2$, the harmonic yield experiences two maxima at the gas-like 1.5:1 ratio, and again in the single-color limit, whereas monolayer hBN shows a monotonic increase as the 2ω component dominates. Combining time-dependent density-functional theory (TDDFT) and a minimal real-space trajectory analysis, we show that these differences arise from the interplay of Bloch velocity and anomalous Hall velocity. The trajectory model quantitatively reproduces the ab-initio results, and offers an intuitive prediction of the harmonics yield without further heavy computation. These insights provide practical guidance for tailoring solid-state HHG and for selecting 2D compounds with desirable responses.

# INTRODUCTION

In gases, the bichromatic circularly-polarized pulses, most commonly counter-rotating bicircular (CRB) ω–2ω pairs (Fig. 1a), drive high-order harmonic generation (HHG) that peaks at a universal "magic" amplitude ratio of 1.5 : 1 [1–4], a result exploited for circular dichroism and attosecond EUV sources [1–11]. Whether this rule survives in solids is unknown. Solid-state HHG promises chip-scale, high-efficiency emitters and direct access to band-topological effects [12–16], yet carriers there move in bands and feel periodic crystallographic potential, which often result in intrinsically material-dependent responses. Solid-state experiments have confirmed spin-angular-momentum conservation [17], valley-selective harmonics [18,19], and new avenues for valleytronic XUV sources [20–22], but have not identified the field ratio that maximizes the harmonic yield. Resolving this gap is essential for filtering candidate materials for compact EUV devices and for using HHG as a probe of electronic topology.

In this work, we select the prototype 2-D monolayers molybdenum disulfide ($MoS_2$) and hexagonal boron nitride (hBN), and investigated their CRB-driven HHG with real-time time-dependent density-functional theory (rt-TDDFT) under infrared femtosecond (fs) pulses. Our simulations show that the gas-phase "magic" ratio breaks down: $MoS_2$ exhibits two yield maxima (at 1.5 : 1 and in the 2ω-only limit), whereas hBN yield increases monotonically with the 2ω component (Fig. 1d). A fully intraband Bloch-oscillation model fails to reproduce these results. To reveal the underlying mechanism, we introduce a real-space model to describe the electron trajectory in crystal following its Bloch velocity and anomalous Hall velocity. Propagating these velocities under the CRB field yields two-dimensional paths in real space, whose geometry changes with the $E_{2\omega}:E_\omega$ ratio. This minimal model reproduces the ab-initio results for both compounds in good agreement, and offers an intuitive explanation of the material-dependent yield from the interplay between band anisotropy and Berry curvature. Our results provide a practical approach for simulating and optimizing HHG in solids. Given that the simple, real-space trajectory model needs only readily available band dispersions and Berry curvature maps, and it works well even in a complicated case such as CRB pulses, it therefore serves as a lightweight yet transferable tool for screening candidate materials and heterostructures before expensive ab-initio dynamics and experimental validation.

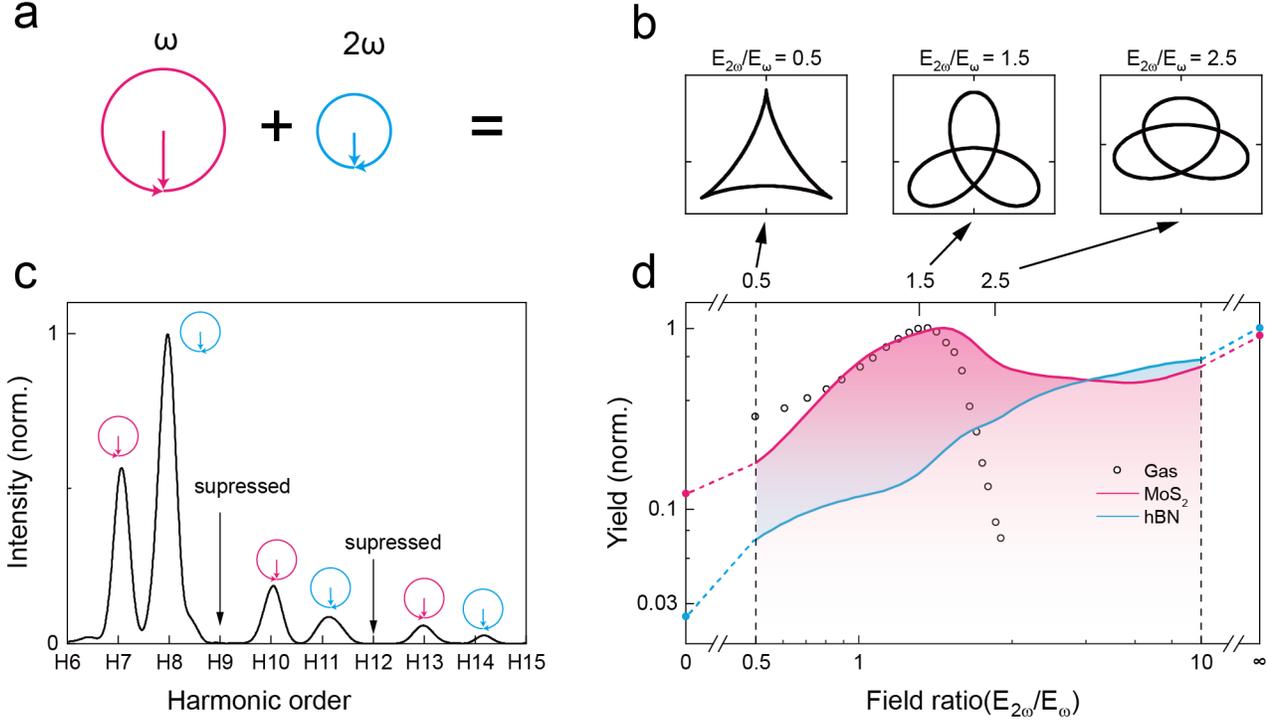

**Figure 1. High-order harmonic generation (HHG) driven by counter-rotating bicircular (CRB) pulses.** (a) Schematic of the ω and 2ω components that have the opposite chirality. (b) Synthesized time-domain electric field under three different $E_{2\omega}:E_\omega$. (c) The chirality of each harmonic and the suppression of 3n-order harmonics in monolayer MoS$_2$ HHG spectrum. (d) Harmonic yields versus $E_{2\omega}:E_\omega$ in Ar gas phase (black hollow circles, experiment, reproduced from [1]), monolayer MoS$_2$ (red line, TDDFT, this work), and monolayer hBN (blue line, TDDFT, this work). Results in the single-color limits ($E_{2\omega}:E_\omega = 0$ and $\infty$) are marked by solid dots. All curves are normalized to their respective maxima.

## RESULTS AND DISCUSSION

All ab-initio calculations were performed with PWMAT [26,27]. Real-time TDDFT simulations were carried out in the velocity gauge in momentum space. The trajectory model was applied based on a semi-classical real-space model [13,24,25,28,29]. The CRB field is a combined laser of left-circularly polarized and right-circularly polarized pulses with an envelope as reported in previous studies [1–8,18–21,30–34]. Details for models and parameters are provided in the Supplemental Material Section S1 and S4.

**Solid-state HHG from real-time TDDFT.**

We first benchmarked our rt-TDDFT implementation against the available linear-polarization data for MoS$_2$

(Fig. 2). We note the small discrepancies here can be due to lack of consideration for dephasing factors and propagation effects [23,24,35–38]. After verifying that the calculated spectral intensities match experiment, we turned to the simulation of CRB pulse. Figure 1c confirms that the calculated solid-state HHG spectrum obey the same spin-angular-momentum selection rule as gases: every 3n-order harmonic is strongly suppressed, in accord with Refs. [7,17]. The field-ratio dependence of the yield, however, departs radically from the gas-phase "magic-ratio" paradigm (Fig. 1d).

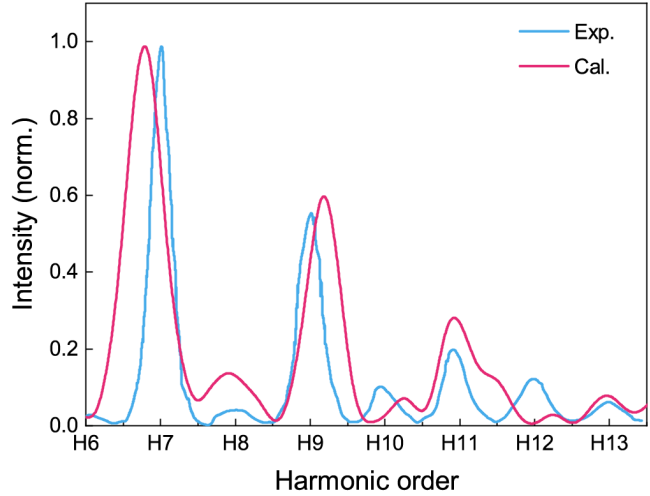

**Figure 2.** Comparison between HHG spectra in monolayer MoS$_2$ under linearly polarized pulse, from TDDFT (red line, this work) and experiment (blue line, reproduced from [39]). The blue and red solid lines represent the HHG of monolayer MoS$_2$ under linearly polarized laser obtained from experiments and computation simulation, respectively. Both curves are normalized to the 7$^{th}$-order harmonic.

For monolayer MoS$_2$, the pulse is with a sine-square envelope and duration of 50 fs, and a center wavelength of ω=0.3 eV (4133 nm) and 2ω=0.6 eV. The total harmonic yield rises with $E_{2\omega}: E_\omega$ increasing, peaks near the canonical 1.5 : 1 ratio, drops, and then recovers to a second maximum at the single-color limit ($E_{2\omega}: E_\omega \to \infty$). For monolayer hBN, the center wavelength is ω=0.775 eV (1600 nm). In contrast to MoS$_2$, its yield grows monotonically over the entire range $0 < E_{2\omega}: E_\omega < \infty$. Neither material shows the sharp decline characteristic of gas-phase HHG when 2ω component dominates. Changing the phase difference between the ω and 2ω components does not affect these trends within numerical accuracy. These contrasting behaviors cannot be explained by the semi-classical three-step model in gas phase [40,41], which attributes the peak at $E_{2\omega}: E_\omega \approx 1.5: 1$ to a universal maximum in the electron–ion recollision probability [1–4]. It indicates that the band structure of the periodic lattice must have non-negligible effects on the harmonic yield under CRB pulses.

**Intraband Bloch-oscillation model.**

To test whether the field-ratio dependence seen in Fig. 1 can be explained in conventional reciprocal-space model, we employed the Bloch oscillation model derived from the semi-classical model [14,39,42,43] in the clean-limit, two-band approximation (Supplemental Material Section S2). In this treatment, the harmonic emission arises from the *intraband* current associated with electronic Bloch oscillation within bands, involving both Bloch velocity and anomalous Hall velocity from Berry curvature. The resulting yields for $MoS_2$, plotted in Fig. 3 along Cartesian $k_x$ and $k_y$ directions (Supplemental Material Section S1), show a single maximum at a lower ratio $E_{2\omega}:E_\omega \approx 1$ and collapse rapidly as the 2ω component increases, in contrast to the rt-TDDFT results. These discrepancies demonstrate that a pure Bloch-oscillation picture is insufficient for CRB-driven solid-state HHG, and the interband contributions is not negligible [13,15,16,25].

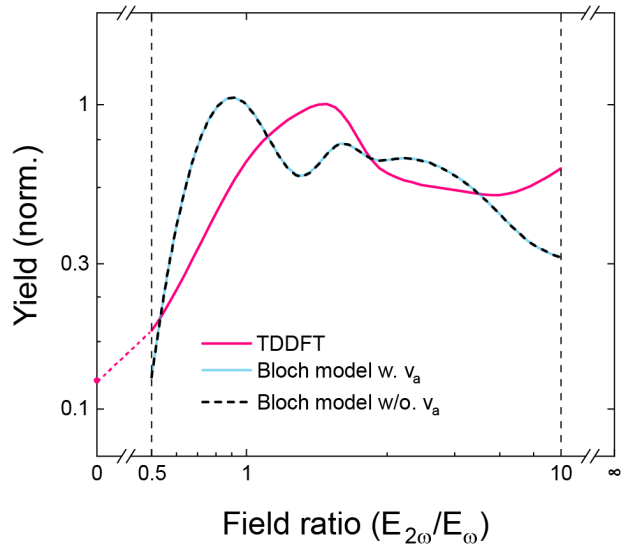

**Figure 3**. **Harmonic yield from an intraband Bloch-oscillation model.** Total harmonic yield versus field ratio $E_{2\omega}:E_\omega$ along $k_x$ (the yield along $k_y$ is identical). Pink solid line: rt-TDDFT benchmark (same data as Fig. 1d). Blue solid line: intraband results from SBE involving both Bloch and anomalous Hall velocities. Black dashed line: intraband results from SBE neglecting the anomalous Hall velocity. All curves are normalized to their respective maxima.

**From momentum-space velocity to real-space trajectory.**

In gases, the celebrated three-step model in real space delivers an intuitive picture that spawned experimental process. Its solid-state analogue, in which the interband emission is tracked through the real-space trajectory of electrons within the periodic lattice potential, has so far been applied only to linearly polarized pulses [25,28,44,45]. Here, we built on that idea and extended it to the two-dimensional case of CRB excitation. In gases, recombination occurs only when the electron returns exactly to the point from which it ionized, i.e., when $r(t) = 0$. Because the ionized electron is near-free, its velocity follows the laser field instantaneously, and previous study showed that a counter-rotating drive with a ratio 1.5:1 maximizes the return probability [1–4]. However, neither prerequisite holds in the solid phase. First, an excited electron can recombine with any lattice sites before it cools down; second, its velocity is dictated by the joint action of the laser field, the band dispersion and, where present, the Berry curvature.

In a one-band approach, the real-space trajectory of electron initially with wave vector $k$ is fully determined once its initial displacement and electron velocity field $v(k, t)$ are known. Under a bicircular drive this field has two contributions:

$$v(\mathbf{k}, t) = \underbrace{\nabla_{\mathbf{k}} \varepsilon_{\text{CB}}(\mathbf{k}, t)}_{v_g} + \underbrace{e\mathbf{E}(t) \times \mathbf{\Omega}_{\text{CB}}(\mathbf{k}, t)}_{v_a} \tag{1}$$

where $\varepsilon_{\text{CB}}(\mathbf{k}, t)$ is the band energy of conduction band, $\mathbf{\Omega}_{\text{CB}}(\mathbf{k}, t)$ is the Berry curvature, and $\mathbf{E}(t)$ is the laser field. The first term on the right-hand side of Eq. (1) is the derivative of band energy to wave vector, known as Bloch velocity, or group velocity $v_g$, while the second term is the anomalous Hall velocity $v_a$. We obtain both band structure and Berry curvature from DFT results on a dense 84x84 k-mesh and a 60x60 k-mesh with second order interpolation method, respectively. The resulting two-dimensional maps for MoS$_2$ and hBN are plotted in Fig. 4. Note that in our trajectory model only the electron propagates, while the hole is assumed to remain localized at its birth (anion) site. Results show that this simple treatment already reproduces the rt-TDDFT harmonic yields. The contrast between the two materials is significant. Nearby the minimal-direct-gap points K and K', MoS$_2$ shows a relatively isotropic and flat band dispersion, together with strong valley-contrasting Berry curvatures, indicating an isotopically small group velocity and large anomalous velocity. hBN, in contrast to MoS$_2$, exhibits stronger, anisotropic dispersion near K and K', but much weaker Berry curvatures.

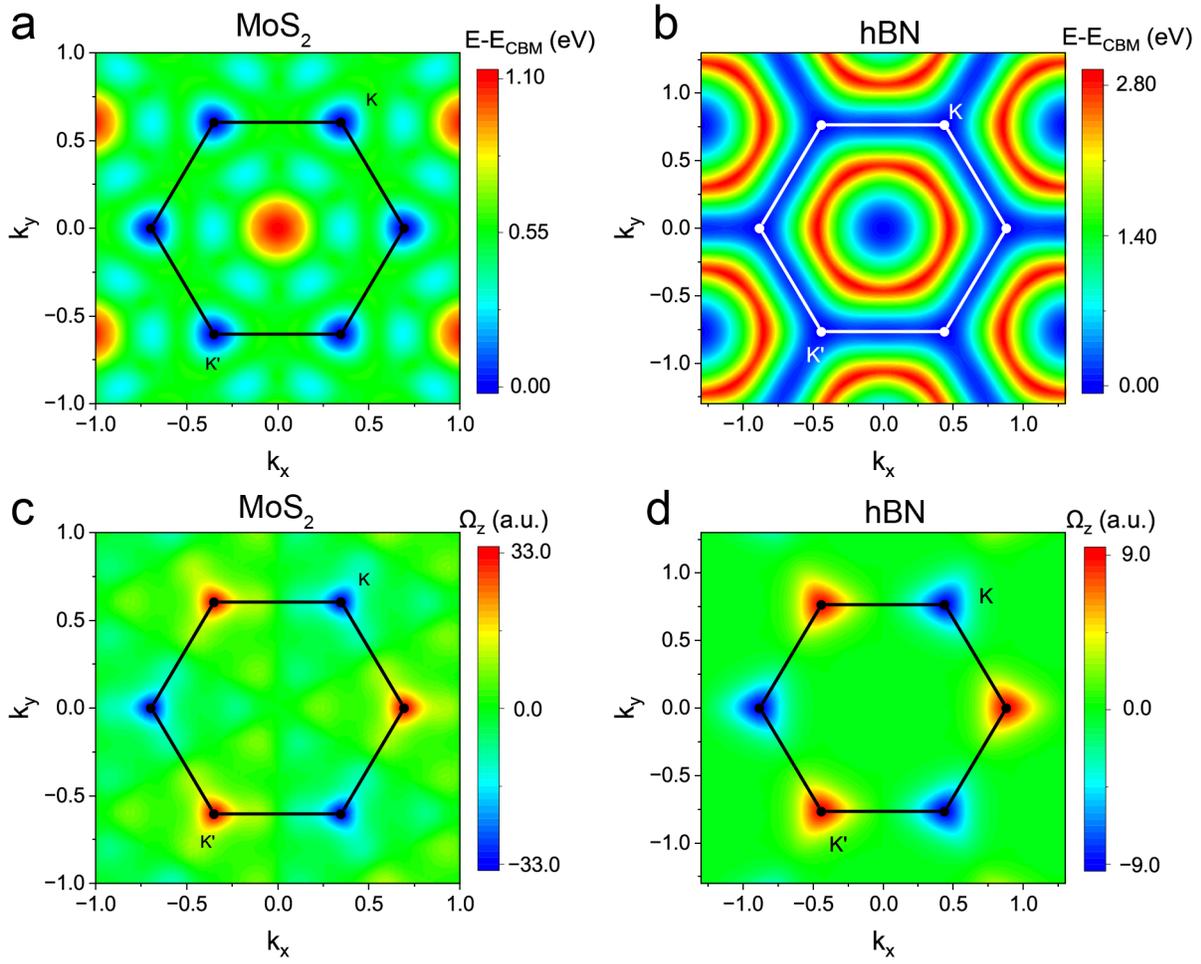

**Figure 4. Band structures and Berry curvatures in momentum space of MoS$_2$ and hBN.** (a) Energy dispersion of conduction-band minimum in MoS$_2$, referenced to the band edge at K. (b) Same quantity for hBN. (c) Out-of-plane Berry curvature of MoS$_2$. Strong, valley-contrasting hotspots appear at K and K'. (d) Berry curvature of hBN, which is much smaller than that in MoS$_2$ near K and K' points. K and K' points, and the boundary of the first Brillouin zone are marked in all plots.

Next, we launch the trajectory at the anion site to mimic the electron right after tunneling from valence band to conduction band. There is only one anion site for each compound due to the space group symmetry. The initial wave vector of the excited electron, however, has two possibilities K and K', as both host the minimal direct gap. Therefore, there are two trajectories $r(K, t)$ and $r(K', t)$ for each compound under each CRB pulse. The yield $Y_{HHG}$ can then be calculated using the following approach [28]

$$Y_{HHG} \propto \int_0^\infty dt \sum_{i \in all\ cation} \left[ e^{-\frac{(r(K,t)-R_i)^2}{r_c^2}} + e^{-\frac{(r(K',t)-R_i)^2}{r_c^2}} \right] \quad (2)$$

where $R_i$ goes through all cation sites, and $r_c$ is the critical radius for recombination. In MoS$_2$ $r_c$ has been taken as the d-valence radius of Mo, 0.73 Å [46], while for hBN it is the s/p-valence radius of B, 0.84 Å [46].

Figure 5 summarizes the key outcome of the trajectory model. Panels 5a and 5b plot the real-space paths of electron that starts at the K valley of MoS$_2$ and hBN, respectively, when the bicircular field has the ratio 1.5 : 1. The corresponding harmonic yields, obtained by adding the K and K' contributions, are compared with the rt-TDDFT benchmarks in Fig. 5c (MoS$_2$) and Fig. 5d (hBN). The agreement is quantitative: in MoS$_2$ the model reproduces the two distinct maxima, one at 1.5 : 1 and another one at single-color limit of 2ω component, while in hBN it correctly gives a yield that grows steadily once the 2ω component dominates. The first MoS$_2$ maximum is captured only when the Berry-curvature-induced anomalous velocity is retained; suppressing that term shifts the peak to field ratio ~ 2 : 1. In hBN where the Berry curvature is smaller, the model predicts a very shallow local maximum at ratio 0.8 : 1. Because the yield rises rapidly beyond that point, the overall curve appears monotonic, matching the ab-initio trend. It confirms that the interplay of group velocity and anomalous velocity can fully explains the distinct bicircular HHG responses of the two monolayers.

Figure 5 also highlights that, even under the same amplitude ratio 1.5 : 1, the real-space motion of the excited electron differs qualitatively between the two monolayers. In MoS$_2$ the trajectory closes into a sharp triangular shape; a "knot" forms at each corner, similar to the bicircular driven field. Once the pulse diminishes, the electron finishes only a minor net drift along the zig-zag axis. In hBN, by contrast, the path evolves into a smooth, rounded triangle that does not follow the laser pattern. There is a noticeably larger accumulated displacement along the same zig-zag direction. These contrasting geometries translate directly into the material-specific harmonic yields shown in Fig. 5c and Fig. 5d.

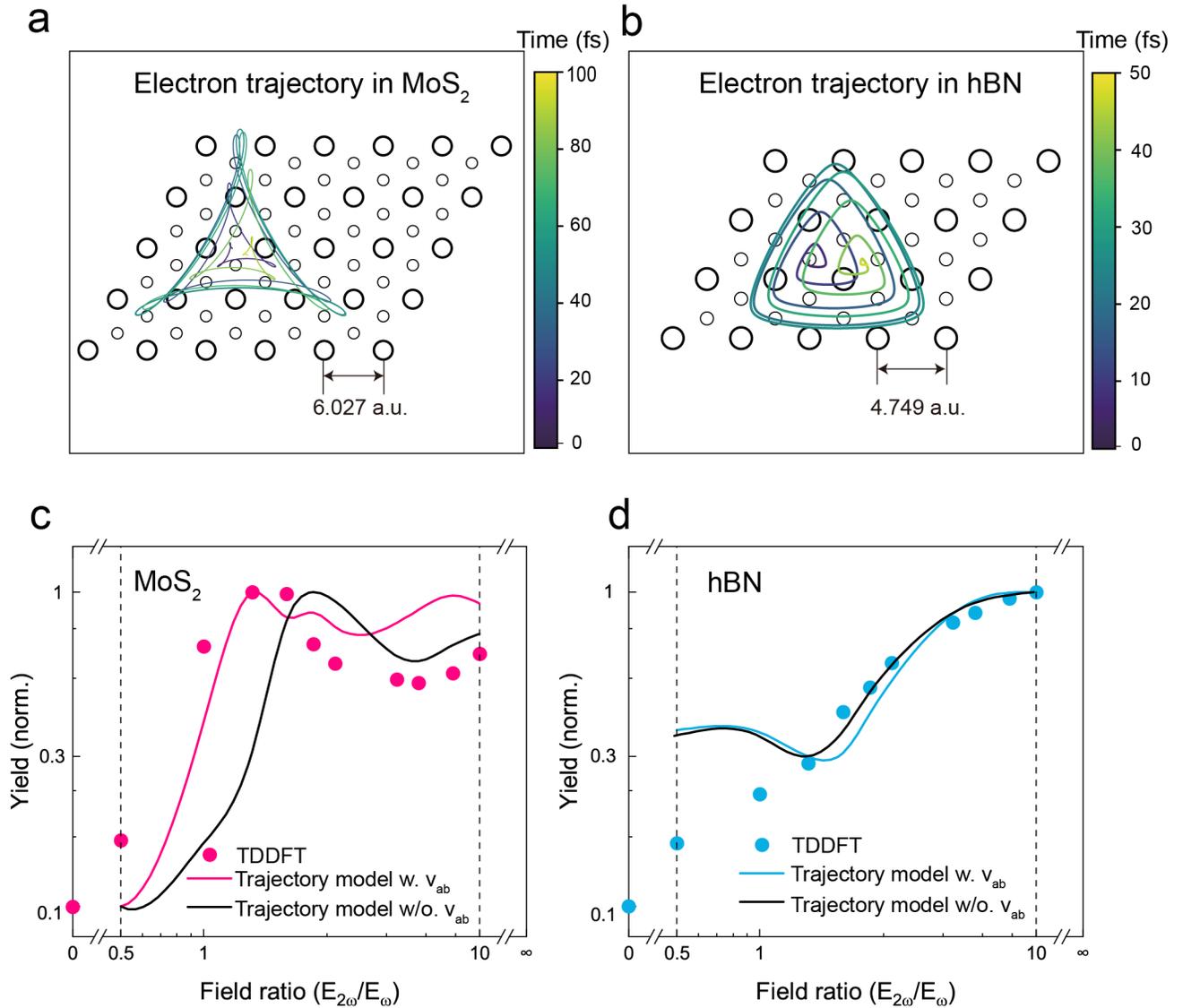

**Figure 5. Real-space electron trajectories and the resulting harmonic yields.** (a) Trajectory of an electron launched from the K valley of monolayer MoS$_2$ under field ratio 1.5 : 1, obtained from the real-space model that includes both the group velocity and the Berry-curvature anomalous velocity. (b) Corresponding trajectory for monolayer hBN. (c) Total harmonic yield of MoS$_2$ versus field ratio (pink circles, rt-TDDFT; pink line, trajectory model with anomalous velocity; black line, trajectory model without anomalous velocity). (d) Total harmonic yield of hBN versus field ratio (blue circles, rt-TDDFT; blue solid line, trajectory model with anomalous velocity; black line, trajectory model without anomalous velocity). In both (c) and (d) the yields are normalized to their respective maxima.

To visualize intuitively how the band structure and Berry curvature reshape the yield, in Fig. 6 we project each real-space trajectory onto the Cartesian x (zigzag) and y (armchair) axes and mark the nearest-neighbor cation positions, which serve as the most probable recombination sites.

For $MoS_2$, the trajectory forms a knotted, sharp-cornered triangle following the shape of the driven field. It is due to the fact that near the K and K' valleys the conduction band is nearly isotropic (Fig. 4a), so the magnitude of the group velocity is weakly correlated with its direction. The projected motions therefore oscillate several times within one optical cycle along both x and y directions (Fig. 6a,c). As the field ratio increases, the "body" of triangle path (Fig. 6a,c, the largest oscillation in one laser cycle) circles monotonically closer to the nearest neighbors, which favors a steadily rising yield, however the "knots" of the path (Fig. 6a,c, the smaller oscillations in one laser cycle) approaches the neighbor at the ratio of 1.5 : 1, moves away from those sites around ratio 3 : 1, then approaches it again when the 2ω component dominates. The superposition of these two tendencies produces the double-peak structure seen in $MoS_2$. A net drift along zigzag direction is almost cancelled because the Berry-curvature-induced anomalous velocity opposes the group velocity drift [1]. In hBN the situation is reversed. Strong in-plane anisotropy of the dispersion (Fig. 4b) turns the trajectory into a rounded loop which executes a single oscillation per cycle. The large group velocity and weak anomalous velocity together result in a sizeable net displacement along the zigzag direction (Fig. 5b and Fig. 6b). As field ratio grows, the projections in both directions are drawn progressively closer to the nearest-neighbor sites (Fig. 6b,d), so the recombination probability, and hence the harmonic yield, rises monotonically.

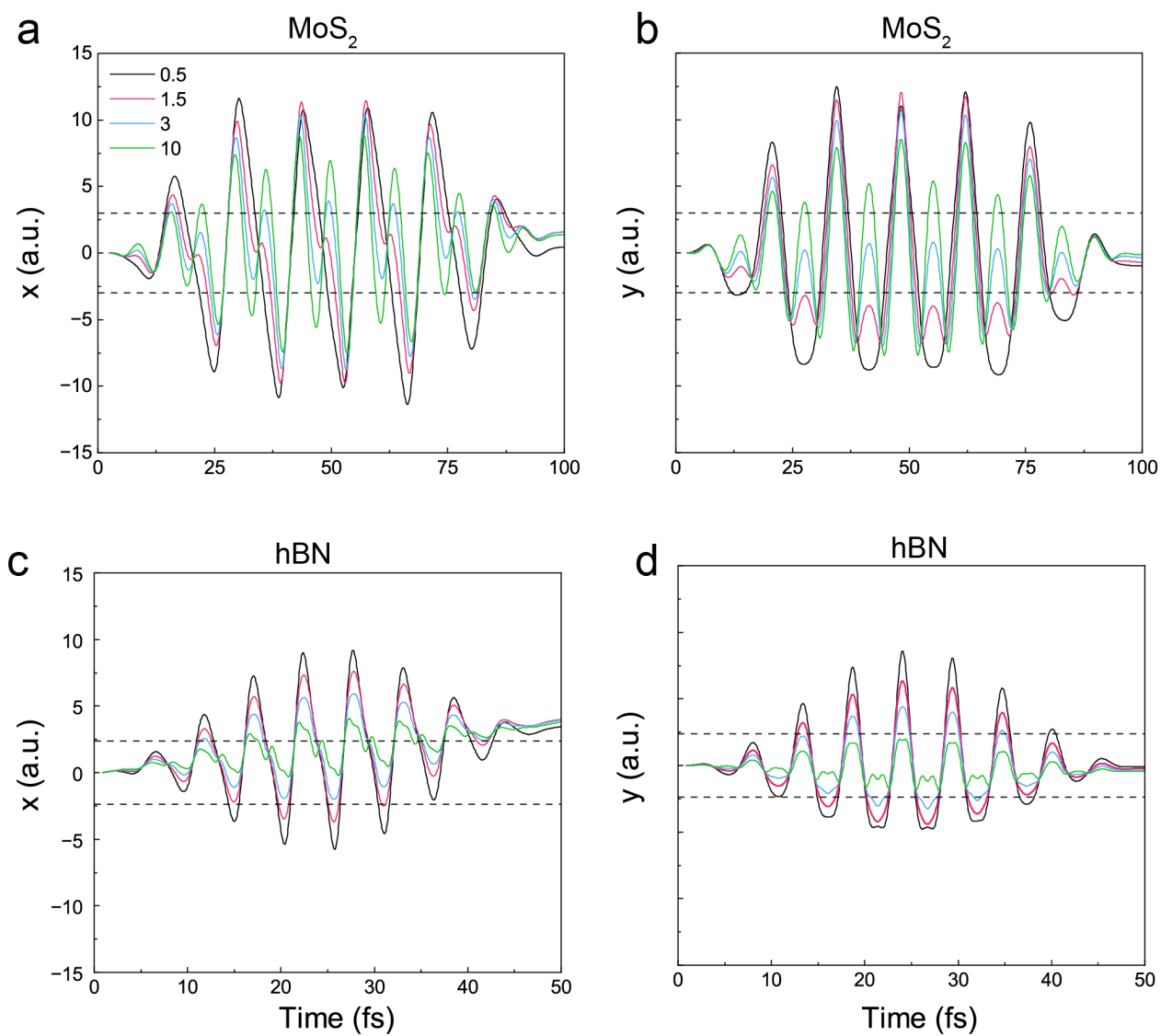

**Figure 6. Cartesian projections of real-space electron trajectories.** (a) x-projection (zigzag direction) of the MoS$_2$ trajectory launched from the K valley; (b) y-projection (armchair direction) of the same MoS$_2$ trajectory; (c) x-projection for hBN; (d) y-projection for hBN. Black, red, blue and green lines correspond to amplitude ratios of 0.5, 1.5, 3.0, and 10, respectively. Horizontal dash lines denote the nearest-neighbor lattice sites. MoS$_2$ exhibits multiple oscillations with small net drift, while hBN shows a single-oscillation behavior and a pronounced drift along zigzag direction, underpinning the double-peak versus monotonic yields discussed in the text.

## CONCLUSION

We have shown that the high-order harmonic generation driven by counter-rotating bicircular fields is intrinsically material-dependent in two-dimensional semiconductors, sharply distinct from the observations in gas phase. The relations of harmonic yield versus field ratio between the 2ω-ω components have been predicted using our real-time TDDFT method in two prototype materials $MoS_2$ and hBN. In monolayer $MoS_2$ the interplay of an almost isotropic band dispersion and strong, valley-contrasting Berry curvature creates two peaks in the yield. In hBN, whose Berry curvature is smaller but band dispersion is more pronounced and highly anisotropic, the yield instead grows monotonically with the 2ω component. A minimal real-space trajectory approach can well explain the yield behavior and show good agreement with the expensive TDDFT calculations. These insights highlight a transferable framework for screening bicircular HHG efficiency with low computational cost, that the harmonic response can be estimated directly from ground-state band structure and Berry curvature, without further TDDFT propagation. While the single-active-electron approach suffices for the two prototypes examined here, extending the model to include multiband and excitonic effects might be necessary for materials with strong correlations or narrow gaps. Experimentally, the predicted material dependence can be tested with current mid-IR systems and exfoliated flakes, opening a direct route toward compact, valley-polarized light sources and band-topology metrology.